\documentclass[pra,onecolumn,superscriptaddress,longbibliography]{revtex4}   

\usepackage[a4paper, total={6.5in, 10 in}]{geometry}
\usepackage{amsbsy,amssymb,amsmath,bm,bbold} 
\usepackage{graphicx,color,epsfig,rotate} 
\usepackage{fancyhdr} 
\usepackage{epstopdf}
\usepackage{float}
\usepackage[colorlinks=true,linkcolor=blue,citecolor=blue,urlcolor=blue]{hyperref}

\usepackage{tikz}

\usepackage{array, booktabs, multirow}
\usepackage{tabularx}

\usepackage[pagewise]{lineno}

\def\bbbc{{\mathchoice {\setbox0=\hbox{$\displaystyle\rm C$}\hbox{\hbox 
to0pt{\kern0.4\wd0\vrule height0.9\ht0\hss}\box0}} 
{\setbox0=\hbox{$\textstyle\rm C$}\hbox{\hbox 
to0pt{\kern0.4\wd0\vrule height0.9\ht0\hss}\box0}} 
{\setbox0=\hbox{$\scriptstyle\rm C$}\hbox{\hbox 
to0pt{\kern0.4\wd0\vrule height0.9\ht0\hss}\box0}} 
{\setbox0=\hbox{$\scriptscriptstyle\rm C$}\hbox{\hbox 
to0pt{\kern0.4\wd0\vrule height0.9\ht0\hss}\box0}}}}

\DeclareMathAlphabet\mathbfcal{OMS}{cmsy}{b}{n}

\AtBeginDocument{%
	\newwrite\bibnotes
	\def\bibnotesext{Notes.bib}
	\immediate\openout\bibnotes=\jobname\bibnotesext
	\immediate\write\bibnotes{@CONTROL{REVTEX41Control}}
	\immediate\write\bibnotes{@CONTROL{%
			apsrev41Control,author="08",editor="1",pages="1",title="0",year="1"}}
	\if@filesw
	\immediate\write\@auxout{\string\citation{apsrev41Control}}%
	\fi
}%

\newcolumntype{P}[1]{>{\centering\arraybackslash}p{#1}}  
\begin{document} 
\title{Parametrically encircled higher-order exceptional points in anti-parity-time symmetric optical microcavities}
\author{Dinesh Beniwal}
\affiliation{School of Cancer \& Pharmaceutical Sciences, King’s College London, London SE1 1UL, United Kingdom}
\author{Arnab Laha}
\author{Adam Miranowicz}
\affiliation{Institute of Spintronics and Quantum Information, Faculty of Physics, Adam Mickiewicz University, 61-614 Pozna\'n, Poland}
\author{Somnath Ghosh}
\email{somiit@rediffmail.com}
\affiliation{Department of Physics, Indian Institute of Technology Jodhpur, Rajasthan-342037, India}

\vspace{2cm}

\begin{abstract}	
The fascinating realm of non-Hermitian physics with the interplay of parity (P) and time-reversal (T) symmetry has been witnessing immense attention in exploring unconventional physics at Exceptional Point (EP) singularities.
Particularly, the physics of PT-symmetry, anti-PT (APT)-symmetry, and the emergence of EPs have ignited fervor in photonics.
Beyond the conventional relation between EP and PT-symmetric phase transitions, this study delves into hosting higher-order EPs in a specially designed APT-symmetric Fabry-P\'erot-type microcavity. 
We unveil the captivating physics of the parametric encirclement schemes to explore the branch-point behaviors of EPs up to order three in terms of successive state-flipping, while optimizing the designed cavity under APT-symmetric constraints.
The insights from our findings are poised to boost research in optical metamaterials, meeting the demands of APT-symmetry and paving the way for a novel class of photonic devices.
\end{abstract}  
\maketitle %

\section{Introduction}

The integration of non-Hermitian physics into photonic systems has garnered significant interest, driven by the ubiquitous presence of loss and gain. Non-Hermitian systems typically exhibit complex eigenvalues. However, the distinctive aspect of a particular class of non-Hermitian systems showing parity-time (PT)-symmetry lies in their possession of real eigenvalues in a specific phase. Owing to the equivalence of the Schr\"odinger equation in quantum mechanics with the Helmholtz equation governing the evolution of an electric field, the exploration of PT-symmetry has gained significant attention, particularly within the realm of wave propagation in optical systems with gain and loss \cite{Ozdemir19,Miri19}. In a PT-symmetric system, the underlying Hamiltonian (say, $H_{\text{PT}}$) adheres to the commutation relation $[\text{PT}, H_{\text{PT}}] = 0$, given that $\text{PT}:\{x,t,i\}\rightarrow\{-x,-t,-i\}$. In the context of an optical system, where the complex potential is represented by the complex refractive index profile $n(x)$, the adherence to PT-symmetry is contingent upon the condition $n(x)=n^*(-x)$. This requirement results in the real part of the refractive index behaving as an even function, while the imaginary part exhibits characteristics of an odd function; i.e., $n_{\text{R}}(x)=n_{\text{R}}(-x)$ and $n_{\text{I}}(x)=-n_{\text{I}}(-x)$ with $n(x)=n_{\text{R}}(x)+in_{\text{I}}(x)$. In this context, the variable $n_{\text{I}}$ is fundamentally tied to the interplay of gain and loss within an optical system. To adhere to PT-symmetry, the distinctive odd function behavior of $n_{\text{I}}$ demands a balanced and symmetrical distribution of gain and loss across a structured background index profile. Therefore, PT-symmetry emerges as a pivotal tool in precisely adjusting gain and loss in photonic devices \cite{Ozdemir19,Miri19}.

A fascinating phase-transition behavior inherent in PT-symmetric systems can be realized through judicious manipulation of control parameters. Such a phase transition from an unbroken PT-phase (characterized by real eigenvalues) to a broken PT-phase (characterized by complex eigenvalues) showcases the emergence of an exceptional point (EP) singularity, which is one of the intriguing non-Hermitian features \cite{Ozdemir19,Miri19}. An EP of the order $n$ (say, EP$n$) is encountered as a topological defect in the system's parameter plane when $n$ number of underlying eigenvalues and their corresponding eigenvectors coalesce simultaneously \cite{Heiss12JPA}. Distinctive avoided-crossing type state interactions can be observed in proximity to an EP, which is pivotal for identifying and characterizing these singularities \cite{Heiss12JPA}. Explicit branch point behavior of an EP is comprehensible through quasistatic state-exchange interactions, governed by encircling the respective EP adiabatically in the system's parameter space \cite{Dembowski04,Laha21EP4}. Intriguingly, the concurrent influence of $(n-1)$ EP2s can mimic the branch-point behavior inherent to an EP$n$, showcasing the complexity of such phenomena \cite{Muller08}. The diverse scientific and technological influence of EPs at the forefront of ongoing research in the fields of photonics and quantum optics facilitate a versatile range of intriguing applications \cite{Ozdemir19,Miri19}, such as controlled lasing with asymmetric state-switching \cite{Laha18,Arkhipov2023}, antilasing \cite{Wang21CPA}, slow-light engineering \cite{Goldzak18}, enhanced nonreciprocity \cite{Laha20}, ultrasensitive detection \cite{Wiersig20}, quantum state tomography \cite{Naghiloo2019}, and advanced quantum state engineering \cite{Minganti2019}. 

Beyond the conventional association between PT-symmetry and EPs, as mentioned above, anti-PT (APT)-symmetry has recently drawn enormous attention while addressing photonic systems with artificial materials \cite{Ge13APT}. APT-symmetry endorses an anti-commutation relation between the PT operator and the fundamental Hamiltonian (say, $H_{\text{APT}}$), i.e., $\{\text{PT}, H_{\text{APT}}\}=0$. Here, $H_{\text{PT}}$ and $H_{\text{APT}}$ adheres an inherent relation $H_{\text{APT}}=\pm iH_{\text{PT}}$. Now, while considering an APT-symmetric optical system, the associated complex potential $n(x)$ adheres to the relationship $n(x)=-n^*(-x)$, where $n_{\text{R}}(x)=-n_{\text{R}}(-x)$ and $n_{\text{I}}(x)=n_{\text{I}}(-x)$. This signifies a deviation from the PT-symmetric case, where $n_{\text{R}}$ and $n_{\text{I}}$ exhibit the characteristics of odd and even functions, respectively. Notably, the odd function characteristics of $n_R$ introduce an additional condition, mandating the use of negative-indexed background materials alongside a precisely balanced gain-loss distribution. The realization of these negative-indexed materials is achievable through metamaterials, characterized by negative permittivity and/or negative permeability. This avenue, enriched with the attributes of APT-symmetry, has recently garnered substantial interest, proving influential in the advancement of artificial photonic components \cite{Zhang2019APTEP,Feng22APTEP,Qi2021APTEP}.

Venturing beyond the conventional connection of PT-symmetry and an EP through a phase transition, the integration of EPs within an APT-symmetric system holds substantial promise to unveil novel perspectives in understanding the EP-induced light dynamics, particularly influenced by the distinctive attributes of negative-indexed materials. While certain studies have shed light on this aspect \cite{Zhang2019APTEP,Feng22APTEP,Qi2021APTEP}, further in-depth research and development is certainly required, especially in the context of encircled higher-order EPs. Here, we report a specially configured gain-loss assisted APT-symmetric optical microcavity to host parametrically encircled EPs up to order three. We exclusively investigate the second-order and third-order branch point behavior of embedded EPs under the APT-symmetric constraints. The intertwining aspects of anti-PT-symmetry and exceptional points not only deepen the understanding of fundamental non-Hermitian physics, but also unlock possibilities for designing novel devices with tailored functionalities, spanning the fields of photonics and quantum optics.
\begin{figure}[b!]
	\centering
	{\includegraphics[width=9cm]{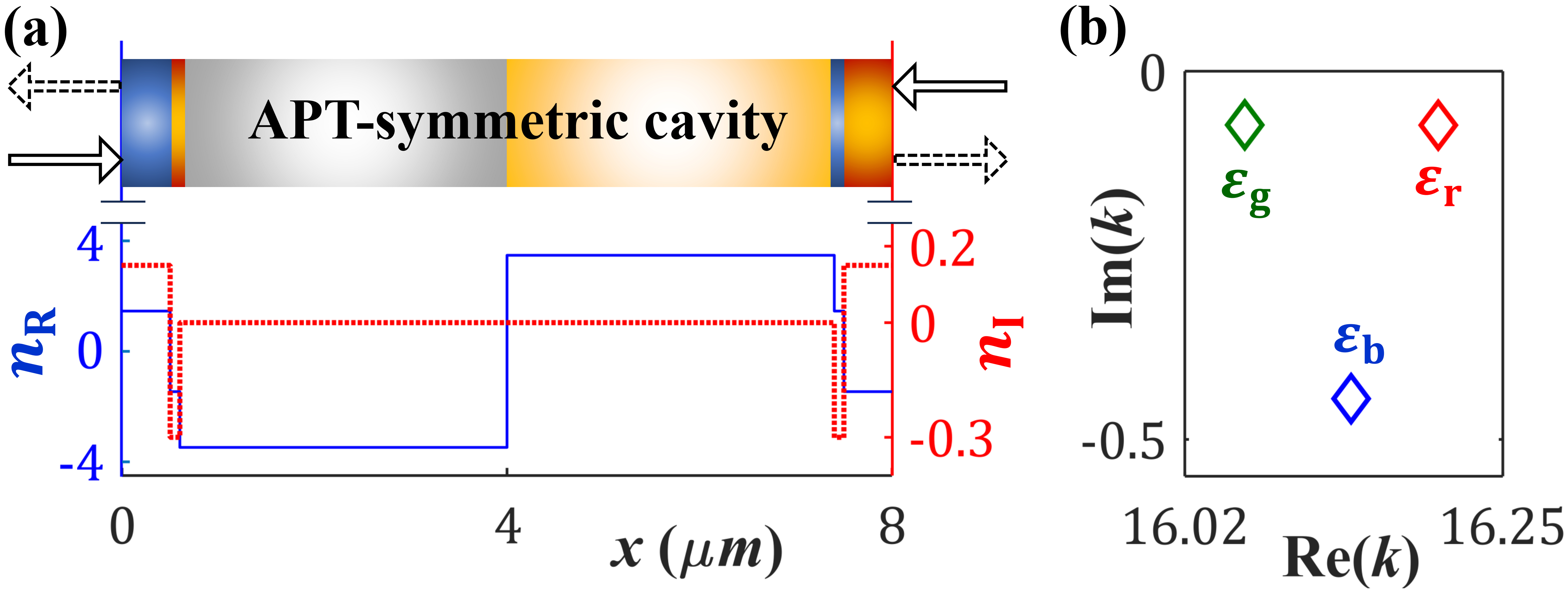}}
	\caption{\textbf{(a)} A schematic representation of the proposed microcavity system exhibiting APT symmetry, accompanied by the corresponding complex refractive index profile $n(x)$. The variations of the $n_{\text{R}}(x)$ and $n_{\text{I}}(x)$ are illustrated by solid blue and dotted red lines, respectively (labeled along the left and right $y$-axes). \textbf{(b)} The coordinates of three chosen $S$-matrix poles, indicated as $\varepsilon_r$, $\varepsilon_b$, and $\varepsilon_g$, situated within the complex $k$-plane, while considering the passive cavity with $\gamma=0$.}
	\label{fig1}
\end{figure}
 
\section{Results and discussions}

\subsection{Cavity configuration to host EPs of different orders}

We engineer a 1D Fabry-Pérot type optical microcavity with a blend of positive and negative indexed background materials, featuring a length of $L=8\mu m$ $(0\le x\le L)$. Here, non-Hermiticity is achieved through a tailored gain-loss profile. The overall refractive index profile $n(x)$ operates according to the functional form:
\begin{equation}
	n(x)=\left\{ 
	\begin{array}{l}
		\pm\,n_1\\ \pm\,n_2-i\gamma\\ \mp\,n_2+i\tau\gamma
	\end{array}\right.
	\begin{array}{l}
		\,:|x-l_0|\in[0,\,l_1],\\\,:|x-l_0|\in[l_1,\,l_2],\\\,:|x-l_0|\in[l_2,\,l_3].
	\end{array}
	\label{nx} 
\end{equation}
In this configuration, the overall cavity system consists of six layers distinguished by two passive index parameters, $\pm n_1=3.48$ and $\pm n_2=1.46$, alongside two gain-loss control parameters, $\gamma$ and $\tau$. Here, $\gamma$ denotes a gain-loss coefficient, while $\tau$ represents a loss-to-gain ratio. The entire setup is designed to maintain APT-symmetry, which can be understood by the distributions of $n_{\text{R}}(x)$  and $n_{\text{I}}(x)$ based on precisely chosen length parameters $l_j$ $(j=0\,\text{---}\,3;\,l_j<L)$ with $l_0=4\mu m=l_3$, $l_1=3.4\mu m$, and $l_2=3.5\mu m$. Figure \ref{fig1}(a) shows a schematic design of the entire cavity system, accompanied by the chosen profile of complex $n(x)$, where the corresponding $n_{\text{R}}(x)$  and $n_{\text{I}}(x)$ follow the features of odd and even functions, respectively. The chosen configuration allows us to uphold APT-symmetry consistently for any specified values of $\gamma$ and $\tau$ throughout our investigation. 

Here, we manifest the physical eigenvalues in terms of the resonance states of the designed microcavity, which are numerically estimated by the scattering ($S$) matrix formalism \cite{Laha21EP4,Ge11}. Consequently, we formulate the $S$-matrix equation as 
\begin{equation}
	\begin{pmatrix} \psi_4 \\ \psi_2 \end{pmatrix}=
	\begin{pmatrix} S_{11} & S_{12} \\ S_{21} & S_{22} \end{pmatrix} 
	\begin{pmatrix} \psi_1 \\ \psi_3 \end{pmatrix},
	\label{smatrix}
\end{equation}
where $\psi_1\,(\psi_3)$ and $\psi_4\,(\psi_2)$ represent the incident and scattered waves at the left (right) side of the cavity. Based on the scattering theory of electromagnetism, the matrix elements $S_{ij}\{k,n(x)\}$ are derived analytically as functions of both frequency $(k)$ and the chosen $n(x)$. Adhering to the energy conservation and causality conditions, the complex poles of the $S$-matrix, residing in the fourth quadrant of the complex $k$-plane with $\text{Re}(k)=m\pi/n_{\text{R}}L$ (where $m$ denotes the order of the poles; $m=1,2,3\ldots$), signify the physical resonance states within the cavity \cite{Laha21EP4}. These poles are determined by solving the equation $1/\max|\text{eig}(S)|=0$ through a numerical root-finding method. We meticulously choose a particular frequency range $16.02\le k \le16.25$ (in $\mu m^{-1}$), encompassing three poles (eigenvalues), denoted as $\varepsilon_r$, $\varepsilon_b$, and $\varepsilon_g$, indicated by three diamond markers of red, blue, and green colors, respectively, in Fig. \ref{fig1}(b). These poles are initially (i.e., when $\gamma=0$) distributed in a nonlinear pattern in the complex $k$-plane, which are mutually coupled with the onset of non-Hermticity (i.e., when $\gamma\ne0$), with a controlled adjustment of gain-loss parameters, $\gamma$ and $\tau$, we delve into their interactions exhibiting avoided-crossing characteristics in the proximity of two second-order branch points.
\begin{figure}[b!]
	\centering
	{\includegraphics[width=9cm]{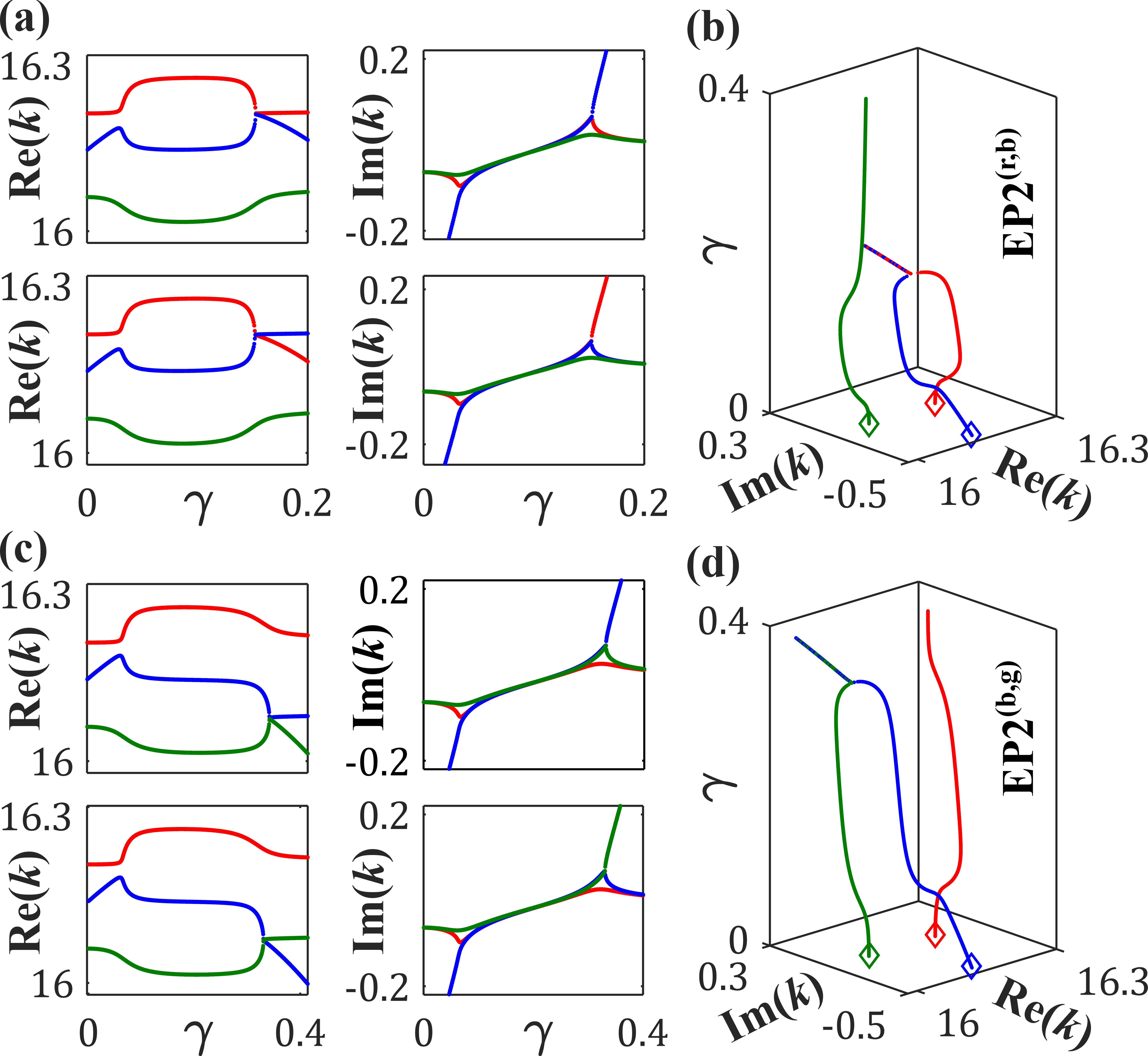}}
	\caption{Trajectories of $\varepsilon_r$, $\varepsilon_b$, and $\varepsilon_g$ (depicted by dotted red, blue, and green curves) with an increasing $\gamma$, while considering different $\tau$-values. \textbf{(a)} For $\tau=0.879$: (upper Panel) An anticrossing and a crossing in Re($k$) and Im($k$), respectively, associated with $\varepsilon_r$ and $\varepsilon_b$; for $\tau=0.883$: (lower Panel) A crossing and an crossing in Re($k$) and Im($k$), respectively, associated with $\varepsilon_r$ and $\varepsilon_b$. \textbf{(b)} At $\tau=0.881$: Emergence of EP2$^{\text{(r,b)}}$ due to the coalescence of $\varepsilon_r$ and $\varepsilon_b$ at $\gamma=0.153$. $\varepsilon_g$ remains away from the strong interaction regime of $\varepsilon_r$ and $\varepsilon_b$ in (a) and (b). \textbf{(c)} For $\tau=0.275$: (upper Panel) An anticrossing and a crossing in Re($k$) and Im($k$), respectively, associated with $\varepsilon_b$ and $\varepsilon_g$; for $\tau=0.279$: (lower Panel) A crossing and an anticrossing in Re($k$) and Im($k$), respectively, associated with $\varepsilon_b$ and $\varepsilon_g$. \textbf{(d)} At $\tau=0.277$: Emergence of EP2$^{\text{(b,g)}}$ due to the coalescence of $\varepsilon_b$ and $\varepsilon_g$ at $\gamma=0.331$. $\varepsilon_r$ remains away from the strong interaction regime of $\varepsilon_b$ and $\varepsilon_g$ in (c) and (d). In (b) and (d), the diamond markers show the locations of $\varepsilon_r$, $\varepsilon_b$, and $\varepsilon_g$ at $\gamma=0$. The unit of $k$ is $\mu m^{-1}$.}
	\label{fig2}
\end{figure}

We monitor the trajectories of $\varepsilon_r$, $\varepsilon_b$, and $\varepsilon_g$ in Fig. \ref{fig2}, while deviating from the passive condition through a gradual increase of $\gamma$, across various $\tau$-values. In Fig. \ref{fig2}(a), we depict two different topological configurations of avoided-crossings among $\varepsilon_r$ and $\varepsilon_b$ in the complex $k$-plane, where $\varepsilon_g$ deviates from the interaction regime. While considering $\tau=0.879$, Re$(k)$ associated with $\varepsilon_r$ and $\varepsilon_b$ undergo an anticrossing, and the corresponding Im$(k)$-values exhibit a crossing with an increasing $\gamma$ (as shown in the upper panel). However, a slight increase in $\tau$ to 0.883 unfolds an exactly opposite topological scenario (for the same variation of $\gamma$) with a crossing and an anticrossing in Re$(k)$ and Im$(k)$, respectively, linked with $\varepsilon_r$ and $\varepsilon_b$ (as shown in the lower panel). Such a topological transition utterly validates the occurrence of a singularity, categorically an EP2, say EP2$^{\text{(r,b)}}$, as illustrated in  Fig. \ref{fig2}(b), where for an intermediary $\tau=0.881$, $\varepsilon_r$ and $\varepsilon_b$ coalesce at $\gamma\approx0.153$, while leaving $\varepsilon_g$ unaffected. In a similar way, two topologically different avoided-crossings among $\varepsilon_b$ and $\varepsilon_g$ with an anticrossing (a crossing) in Re$(k)$ and a crossing (an anticrossing) in Im$(k)$ can be observed in the upper panel (lower panel) of Fig. \ref{fig2}(c), while varying $\gamma$ for a chosen $\tau=0.275$ ($\tau=0.279$). This implies the emergence of another EP2, say EP2$^{\text{(b,g)}}$, as shown in Fig. \ref{fig2}(d), where $\varepsilon_b$ and $\varepsilon_g$ coalesce at $\gamma\approx0.331$ for an intermediary $\tau=0.277$, keeping $\varepsilon_r$ unaffected.

Therefore, we observe a unique scenario involving the three chosen poles, where $\varepsilon_b$ becomes analytically connected to $\varepsilon_r$ and $\varepsilon_g$ through two interconnected EP2s, i.e., EP2$^{\text{(r,b)}}$ and EP2$^{\text{(b,g)}}$, positioned at coordinates (0.153, 0.881) and (0.331, 0.277), respectively, within the $(\gamma,\tau)$-plane (2D parameter space). Notably, while a specific pair of poles coalesce at an EP2, the third pole remains unaffected. Such an intricate scenario occurring within a particular interaction regime leads to the emergence of a third-order branch point, specifically an EP3, where all three interacting poles are intricately linked \cite{Muller08,Laha21EP4}. Here, a particular APT-symmetric phase persists across the entire adjustable range of $\gamma$ and $\tau$ due to the distinctive configuration of the cavity.

\subsection{Effect of different parametric encirclement schemes: Toward exploring second-order and third-order branch point properties}

We investigate the branch-point behaviors of the embedded EP2s by driving the perturbation quasistatically in terms of a patterned gain-loss variation along a closed loop in the $(\gamma,\tau)$-plane, subject to various conditions under the APT-symmetric constraints. The shape of the chosen parametric loop is given by
\begin{equation}
	\gamma(\phi)=\gamma_0\sin(\phi/2)\quad\text{and}\quad\tau(\phi)=\tau_0-p\sin\phi.
	\label{enc}
\end{equation}
Here, $(\gamma_0,\tau_0)$ and $p,\,(<1;\,\ne0)$ represent the characteristic parameters determining the number of EP2s to be encircled ($\gamma_0$ must exceed the $\gamma$-coordinate of the respective EP2 to be encircled), where $p>0\,(p<0)$ characterizes an anticlockwise (a clockwise) encirclement scheme for $0\le\phi\le2\pi$. The chosen shape of the parametric loop ensures that the encirclement process begins ($\phi=0$) and ends ($\phi=2\pi$) at the passive cavity condition ($\gamma=0$).
\begin{figure}[b!]
	\centering
	{\includegraphics[width=9cm]{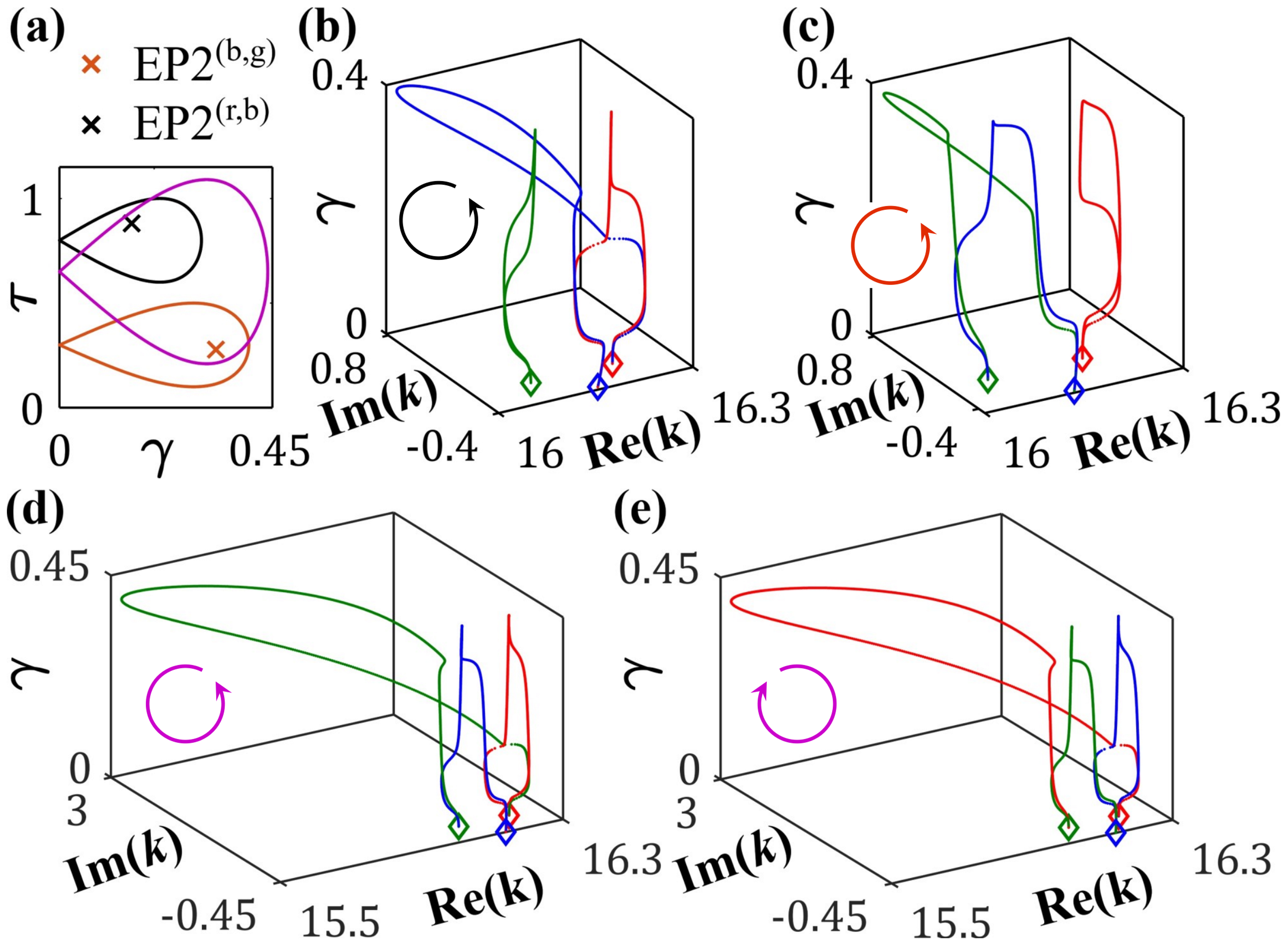}}
	\caption{\textbf{(a)} The coordinates of EP2$^{\text{(r,b)}}$ and EP2$^{\text{(b,g)}}$, along with three chosen quasistatic encirclement schemes. The black and orange loops individually encircle EP2$^{\text{(r,b)}}$ and EP2$^{\text{(b,g)}}$, respectively, whereas the violet loop encircles both the EP2s simultaneously. The trajectories of $\varepsilon_r$, $\varepsilon_b$ and $\varepsilon_g$ (depicted by dotted red, blue, and green curves) in the complex $k$-plane, while considering \textbf{(b)} an anticlockwise variation of $\gamma$ and $\tau$ along the black loop, exhibiting the adiabatic permutations $\varepsilon_r\rightarrow\varepsilon_b\rightarrow\varepsilon_r$ and $\varepsilon_g\rightarrow\varepsilon_g$; \textbf{(c)} an anticlockwise variation of $\gamma$ and $\tau$ along the orange loop, exhibiting the adiabatic permutations $\varepsilon_b\rightarrow\varepsilon_g\rightarrow\varepsilon_b$ and $\varepsilon_r\rightarrow\varepsilon_r$; \textbf{(d)} an anticlockwise variation of $\gamma$ and $\tau$ along the violet loop, exhibiting the adiabatic permutation $\varepsilon_r\rightarrow\varepsilon_b\rightarrow\varepsilon_g\rightarrow\varepsilon_r$; \textbf{(e)} a clockwise variation of $\gamma$ and $\tau$ along the violet loop, exhibiting the adiabatic permutation $\varepsilon_r\rightarrow\varepsilon_g\rightarrow\varepsilon_b\rightarrow\varepsilon_r$. In (b)-(e), the diamond markers show the locations of $\varepsilon_r$, $\varepsilon_b$, and $\varepsilon_g$ at $\phi=0$. The unit of $k$ is $\mu m^{-1}$.}
	\label{fig3}
\end{figure} 

Here, the features of the second-order branch point of the embedded EP2s become evident when encircling them individually. However, an encirclement scheme enclosing both the EP2s simultaneously reveals the nature of a third-order branch point. Figure \ref{fig3}(a) illustrates the coordinates of EP2$^{\text{(r,b)}}$ and EP2$^{\text{(b,g)}}$, alongside three distinct encirclement schemes in the $(\gamma,\tau)$-plane. The encirclement patterns are delineated as follows: a black loop, characterized by $\gamma_0=0.3$, $\tau_0=0.8$, and $p=0.2$, encloses only EP2$^{\text{(r,b)}}$; an orange loop, governed by $\gamma_0=0.4$, $\tau_0=0.3$, and $p=0.2$, encloses EP2$^{\text{(b,g)}}$ solely; whereas a violet loop, characterized by $\gamma_0=0.44$, $\tau_0=0.65$, and $p=0.44$, encircles both the connected EP2s, simultaneously. Our proposed microcavity design guarantees the preservation of APT-symmetry even under applied perturbations that vary along the chosen loops based on \eqref{enc}. The topological effects induced by the chosen encirclement schemes are investigated by tracing the trajectories of $\varepsilon_r$, $\varepsilon_b$, and $\varepsilon_g$, as portrayed in Figs. \ref{fig3}(b)-\ref{fig3}(e). Here, each point of evolution on the trajectory of a specific pole in the complex $k$-plane aligns with a corresponding point of evolution on a specific loop in the $(\gamma,\tau)$-plane. 

Now, while considering an anticlockwise encirclement by varying $\gamma$ and $\tau$ quasistatically along the black loop [that encircles only EP2$^{\text{(r,b)}}$, and keeps EP2$^{\text{(b,g)}}$ outside], the poles $\varepsilon_r$ and $\varepsilon_b$, which are connected through EP2$^{\text{(r,b)}}$, exchanges their initial positions adiabatically in the complex $k$-plane. Upon completing a full $2\pi$ rotation along the loop, $\varepsilon_r$ and $\varepsilon_b$ completely swap their frequencies, transitioning as $\varepsilon_r\rightarrow\varepsilon_b\rightarrow\varepsilon_r$, as depicted in Figure \ref{fig3}(b). Nevertheless, this structured perturbation around EP2$^{\text{(r,b)}}$ does not impact $\varepsilon_g$ [i.e., $\varepsilon_g\rightarrow\varepsilon_g$, as can be observed in Fig. \ref{fig3}(b)], which remains at the same frequency level at the end of the encirclement process. In a similar fashion, a complete $2\pi$ anticlockwise parametric rotation along the orange loop [that encircles only EP2$^{\text{(b,g)}}$, keeping EP2$^{\text{(r,b)}}$ outside] results in an adiabatic frequency-swapping between $\varepsilon_b$ and $\varepsilon_g$ (like, $\varepsilon_b\rightarrow\varepsilon_g\rightarrow\varepsilon_b$), while leaving $\varepsilon_r$ unaffected (i.e., $\varepsilon_r\rightarrow\varepsilon_r$), as illustrated in Figure \ref{fig3}(c). The unconventional interactions, as observed among the three cavity states in Figs. \ref{fig3}(b) and \ref{fig3}(c), showcasing distinct state-flipping characteristics within two corresponding pairs, unfold the individual second-order branch-point behavior of EP2$^{\text{(r,b)}}$ and EP2$^{\text{(b,g)}}$. It is noteworthy that a $2\pi$ clockwise rotation of $\gamma$ and $\tau$ along both the black and orange loops results in a similar permutation among the cavity-states (however, two exchanging cavity-states alters their trajectories), conveying the individual chiral property of both the embedded EP2s. Here, to restore the initial frequencies by the chosen cavity states, a complete $4\pi$ rotation is required for the encirclement schemes along both black and orange loops in any of the directions.

To delve into the intriguing properties of an EP3 as a third-order branch point, we consider a quasistatic variation of $\gamma$ and $\tau$ along the violet loop, encompassing both EP2$^{\text{(r,b)}}$ and EP2$^{\text{(b,g)}}$ simultaneously. Such a patterned perturbation interestingly facilitates a topological switching among all three interacting poles interconnected via EP2$^{\text{(r,b)}}$ and EP2$^{\text{(b,g)}}$. Notably, a complete $2\pi$ rotation in the anticlockwise direction results in a successive and adiabatic exchange of frequencies among $\varepsilon_r$, $\varepsilon_b$ and $\varepsilon_g$, following the sequence $\varepsilon_r\rightarrow\varepsilon_b\rightarrow\varepsilon_g\rightarrow\varepsilon_r$ within the complex $k$-plane, as shown in Fig. \ref{fig3}(d). This manifestation vividly showcases the third-order branch point behavior of an EP3 in the presence of interconnected EP2s. Furthermore, the effect of a complete $2\pi$ rotation in the clockwise direction can be distinguished from the sequence of the resulting successive state exchange phenomena, as illustrated in Fig. \ref{fig3}(e). Here, we can observe a successive and adiabatic frequency switching phenomenon such as $\varepsilon_r\rightarrow\varepsilon_g\rightarrow\varepsilon_b\rightarrow\varepsilon_r$, unlike the case for the anticlockwise encirclement process. This disparity underscores a breakdown of the chiral property alongside the presence of a third-order branch point, i.e., an EP3. Such a breakdown of chirality offers a promising avenue for implementing a programmable state-switching mechanism in the proximity of an EP3 (i.e., along the violet loop enclosing both the connected EP2s), as depicted in Table \ref{t1}. This table outlines the required rotations, either clockwise or anticlockwise, for the transition between states. Notably, a full $6\pi$ rotation is necessary to revert to the initial cavity states.

\begin{table}[htpb]
	\caption{Programmable state-switching process [$n\circlearrowleft$ and $n\circlearrowright$ mean $2n\pi$ anticlockwise and clockwise rotations, respectively].}\label{t1}
	\centering
	\begin{tabular}{P{2.5cm}P{2.5cm}P{2.5cm}P{2.5cm}}
		\hline
		\multirow{2}{*}{Initial States}&\multicolumn{3}{c}{Final states}\\ \cline{2-4} 
		&$\varepsilon_r$&$\varepsilon_b$&$\varepsilon_g$\\
		\hline
		$\varepsilon_r$ & 3 $\circlearrowleft$ \space or \space 3 $\circlearrowright$ & 1 $\circlearrowleft$ \space or \space 2 $\circlearrowright$ & 2 $\circlearrowleft$ \space or \space 1 $\circlearrowright$\\
		$\varepsilon_b$ & 2 $\circlearrowleft$ \space or \space 1 $\circlearrowright$ & 3 $\circlearrowleft$ \space or \space 3 $\circlearrowright$ & 1 $\circlearrowleft$ \space or \space 2 $\circlearrowright$\\
		$\varepsilon_g$ & 1 $\circlearrowleft$ \space or \space 2 $\circlearrowright$ & 2 $\circlearrowleft$ \space or \space 1 $\circlearrowright$ & 3 $\circlearrowleft$ \space or \space 3 $\circlearrowright$\\
		\hline
	\end{tabular}
\end{table}

\section{Conclusion}

In conclusion, this research delves into the intricate characteristics of higher-order EPs within a specialty gain-loss assisted optical microcavity adhering to APT-symmetry. Beyond the widely explored connection between EPs and PT-symmetry, the inclusion of APT-symmetry adds a new dimension to the physics dealing with the topological interplay of gain-loss and negative refractive indexed synthetic materials and expands the repertoire of platforms available for manipulating light. We specifically focus on encountering an EP3 with two interconnected EP2s and unfolding the topological properties of second-order and third-order branch points by considering different parametric encirclement schemes. A successive and adiabatic switching process is revealed among up to three cavity states. Our chosen microcavity configuration with a blend of positive and negative indexed background materials and the topology of the encircling loops in the gain-loss parameter space ensures the maintenance of a particular APT-symmetric phase throughout the operation. Furthermore, leveraging the intriguing chiral aspects uncovered, we explore a programmable state-switching scheme as a potential application of the designed APT-symmetric microcavity. These findings contribute significantly to our comprehension of integrating non-Hermitian physics into classical wave-based systems reliant on metamaterials, thereby advancing the development of artificial devices for all-photonic applications.
\\
\\
\noindent\textbf{Funding.} AL and AM acknowledge the financial support from the Maestro Grant of the Polish National Science Center (NCN), Grant No. DEC-2019/34/A/ST2/00081.

\bibliography{ref}

\begin{thebibliography}{19}%
\makeatletter
\providecommand \@ifxundefined [1]{%
 \@ifx{#1\undefined}
}%
\providecommand \@ifnum [1]{%
 \ifnum #1\expandafter \@firstoftwo
 \else \expandafter \@secondoftwo
 \fi
}%
\providecommand \@ifx [1]{%
 \ifx #1\expandafter \@firstoftwo
 \else \expandafter \@secondoftwo
 \fi
}%
\providecommand \natexlab [1]{#1}%
\providecommand \enquote  [1]{``#1''}%
\providecommand \bibnamefont  [1]{#1}%
\providecommand \bibfnamefont [1]{#1}%
\providecommand \citenamefont [1]{#1}%
\providecommand \href@noop [0]{\@secondoftwo}%
\providecommand \href [0]{\begingroup \@sanitize@url \@href}%
\providecommand \@href[1]{\@@startlink{#1}\@@href}%
\providecommand \@@href[1]{\endgroup#1\@@endlink}%
\providecommand \@sanitize@url [0]{\catcode `\\12\catcode `\$12\catcode
  `\&12\catcode `\#12\catcode `\^12\catcode `\_12\catcode `\%12\relax}%
\providecommand \@@startlink[1]{}%
\providecommand \@@endlink[0]{}%
\providecommand \url  [0]{\begingroup\@sanitize@url \@url }%
\providecommand \@url [1]{\endgroup\@href {#1}{\urlprefix }}%
\providecommand \urlprefix  [0]{URL }%
\providecommand \Eprint [0]{\href }%
\providecommand \doibase [0]{http://dx.doi.org/}%
\providecommand \selectlanguage [0]{\@gobble}%
\providecommand \bibinfo  [0]{\@secondoftwo}%
\providecommand \bibfield  [0]{\@secondoftwo}%
\providecommand \translation [1]{[#1]}%
\providecommand \BibitemOpen [0]{}%
\providecommand \bibitemStop [0]{}%
\providecommand \bibitemNoStop [0]{.\EOS\space}%
\providecommand \EOS [0]{\spacefactor3000\relax}%
\providecommand \BibitemShut  [1]{\csname bibitem#1\endcsname}%
\let\auto@bib@innerbib\@empty
\bibitem [{\citenamefont {{\"O}zdemir}\ \emph {et~al.}(2019)\citenamefont
  {{\"O}zdemir}, \citenamefont {Rotter}, \citenamefont {Nori},\ and\
  \citenamefont {Yang}}]{Ozdemir19}%
  \BibitemOpen
  \bibfield  {author} {\bibinfo {author} {\bibfnamefont {{\c{S}}.~K.}\
  \bibnamefont {{\"O}zdemir}}, \bibinfo {author} {\bibfnamefont
  {S.}~\bibnamefont {Rotter}}, \bibinfo {author} {\bibfnamefont
  {F.}~\bibnamefont {Nori}}, \ and\ \bibinfo {author} {\bibfnamefont
  {L.}~\bibnamefont {Yang}},\ }\bibfield  {title} {\enquote {\bibinfo {title}
  {Parity--time symmetry and exceptional points in photonics},}\ }\href
  {\doibase 10.1038/s41563-019-0304-9} {\bibfield  {journal} {\bibinfo
  {journal} {Nat. Mater.}\ }\textbf {\bibinfo {volume} {18}},\ \bibinfo {pages}
  {783--798} (\bibinfo {year} {2019})}\BibitemShut {NoStop}%
\bibitem [{\citenamefont {Miri}\ and\ \citenamefont {Al{\`u}}(2019)}]{Miri19}%
  \BibitemOpen
  \bibfield  {author} {\bibinfo {author} {\bibfnamefont {M.-A.}\ \bibnamefont
  {Miri}}\ and\ \bibinfo {author} {\bibfnamefont {A.}~\bibnamefont {Al{\`u}}},\
  }\bibfield  {title} {\enquote {\bibinfo {title} {Exceptional points in optics
  and photonics},}\ }\href {\doibase 10.1126/science.aar7709} {\bibfield
  {journal} {\bibinfo  {journal} {Science}\ }\textbf {\bibinfo {volume}
  {363}},\ \bibinfo {pages} {eaar7709} (\bibinfo {year} {2019})}\BibitemShut
  {NoStop}%
\bibitem [{\citenamefont {Heiss}(2012)}]{Heiss12JPA}%
  \BibitemOpen
  \bibfield  {author} {\bibinfo {author} {\bibfnamefont {W.~D.}\ \bibnamefont
  {Heiss}},\ }\bibfield  {title} {\enquote {\bibinfo {title} {The physics of
  exceptional points},}\ }\href {\doibase 10.1088/1751-8113/45/44/444016}
  {\bibfield  {journal} {\bibinfo  {journal} {J. Phys. A: Math. Theor.}\
  }\textbf {\bibinfo {volume} {45}},\ \bibinfo {pages} {444016} (\bibinfo
  {year} {2012})}\BibitemShut {NoStop}%
\bibitem [{\citenamefont {Dembowski}\ \emph {et~al.}(2004)\citenamefont
  {Dembowski}, \citenamefont {Dietz}, \citenamefont {Gr\"af}, \citenamefont
  {Harney}, \citenamefont {Heine}, \citenamefont {Heiss},\ and\ \citenamefont
  {Richter}}]{Dembowski04}%
  \BibitemOpen
  \bibfield  {author} {\bibinfo {author} {\bibfnamefont {C.}~\bibnamefont
  {Dembowski}}, \bibinfo {author} {\bibfnamefont {B.}~\bibnamefont {Dietz}},
  \bibinfo {author} {\bibfnamefont {H.-D.}\ \bibnamefont {Gr\"af}}, \bibinfo
  {author} {\bibfnamefont {H.~L.}\ \bibnamefont {Harney}}, \bibinfo {author}
  {\bibfnamefont {A.}~\bibnamefont {Heine}}, \bibinfo {author} {\bibfnamefont
  {W.~D.}\ \bibnamefont {Heiss}}, \ and\ \bibinfo {author} {\bibfnamefont
  {A.}~\bibnamefont {Richter}},\ }\bibfield  {title} {\enquote {\bibinfo
  {title} {Encircling an exceptional point},}\ }\href {\doibase
  10.1103/PhysRevE.69.056216} {\bibfield  {journal} {\bibinfo  {journal} {Phys.
  Rev. E}\ }\textbf {\bibinfo {volume} {69}},\ \bibinfo {pages} {056216}
  (\bibinfo {year} {2004})}\BibitemShut {NoStop}%
\bibitem [{\citenamefont {Laha}\ \emph {et~al.}(2021)\citenamefont {Laha},
  \citenamefont {Beniwal},\ and\ \citenamefont {Ghosh}}]{Laha21EP4}%
  \BibitemOpen
  \bibfield  {author} {\bibinfo {author} {\bibfnamefont {A.}~\bibnamefont
  {Laha}}, \bibinfo {author} {\bibfnamefont {D.}~\bibnamefont {Beniwal}}, \
  and\ \bibinfo {author} {\bibfnamefont {S.}~\bibnamefont {Ghosh}},\ }\bibfield
   {title} {\enquote {\bibinfo {title} {Successive switching among four states
  in a gain-loss-assisted optical microcavity hosting exceptional points up to
  order four},}\ }\href {\doibase 10.1103/PhysRevA.103.023526} {\bibfield
  {journal} {\bibinfo  {journal} {Phys. Rev. A}\ }\textbf {\bibinfo {volume}
  {103}},\ \bibinfo {pages} {023526} (\bibinfo {year} {2021})}\BibitemShut
  {NoStop}%
\bibitem [{\citenamefont {M\"uller}\ and\ \citenamefont
  {Rotter}(2008)}]{Muller08}%
  \BibitemOpen
  \bibfield  {author} {\bibinfo {author} {\bibfnamefont {M.}~\bibnamefont
  {M\"uller}}\ and\ \bibinfo {author} {\bibfnamefont {I.}~\bibnamefont
  {Rotter}},\ }\bibfield  {title} {\enquote {\bibinfo {title} {Exceptional
  points in open quantum systems},}\ }\href {\doibase
  10.1088/1751-8113/41/24/244018} {\bibfield  {journal} {\bibinfo  {journal}
  {J. Phys. A: Math. Theor.}\ }\textbf {\bibinfo {volume} {41}},\ \bibinfo
  {pages} {244018} (\bibinfo {year} {2008})}\BibitemShut {NoStop}%
\bibitem [{\citenamefont {Laha}\ \emph {et~al.}(2018)\citenamefont {Laha},
  \citenamefont {Biswas},\ and\ \citenamefont {Ghosh}}]{Laha18}%
  \BibitemOpen
  \bibfield  {author} {\bibinfo {author} {\bibfnamefont {A.}~\bibnamefont
  {Laha}}, \bibinfo {author} {\bibfnamefont {A.}~\bibnamefont {Biswas}}, \ and\
  \bibinfo {author} {\bibfnamefont {S.}~\bibnamefont {Ghosh}},\ }\bibfield
  {title} {\enquote {\bibinfo {title} {Nonadiabatic modal dynamics around
  exceptional points in an all-lossy dual-mode optical waveguide: Toward
  chirality-driven asymmetric mode conversion},}\ }\href {\doibase
  10.1103/PhysRevApplied.10.054008} {\bibfield  {journal} {\bibinfo  {journal}
  {Phys. Rev. Applied}\ }\textbf {\bibinfo {volume} {10}},\ \bibinfo {pages}
  {054008} (\bibinfo {year} {2018})}\BibitemShut {NoStop}%
\bibitem [{\citenamefont {Arkhipov}\ \emph {et~al.}(2023)\citenamefont
  {Arkhipov}, \citenamefont {Miranowicz}, \citenamefont {Minganti},
  \citenamefont {{\"O}zdemir},\ and\ \citenamefont {Nori}}]{Arkhipov2023}%
  \BibitemOpen
  \bibfield  {author} {\bibinfo {author} {\bibfnamefont {I.~I.}\ \bibnamefont
  {Arkhipov}}, \bibinfo {author} {\bibfnamefont {A.}~\bibnamefont
  {Miranowicz}}, \bibinfo {author} {\bibfnamefont {F.}~\bibnamefont
  {Minganti}}, \bibinfo {author} {\bibfnamefont {{\c{S}}.~K.}\ \bibnamefont
  {{\"O}zdemir}}, \ and\ \bibinfo {author} {\bibfnamefont {F.}~\bibnamefont
  {Nori}},\ }\bibfield  {title} {\enquote {\bibinfo {title} {Dynamically
  crossing diabolic points while encircling exceptional curves: A programmable
  symmetric-asymmetric multimode switch},}\ }\href {\doibase
  10.1038/s41467-023-37275-5} {\bibfield  {journal} {\bibinfo  {journal}
  {Nature Communications}\ }\textbf {\bibinfo {volume} {14}},\ \bibinfo {pages}
  {2076} (\bibinfo {year} {2023})}\BibitemShut {NoStop}%
\bibitem [{\citenamefont {Wang}\ \emph {et~al.}(2021)\citenamefont {Wang},
  \citenamefont {Sweeney}, \citenamefont {Stone},\ and\ \citenamefont
  {Yang}}]{Wang21CPA}%
  \BibitemOpen
  \bibfield  {author} {\bibinfo {author} {\bibfnamefont {C.}~\bibnamefont
  {Wang}}, \bibinfo {author} {\bibfnamefont {W.~R.}\ \bibnamefont {Sweeney}},
  \bibinfo {author} {\bibfnamefont {A.~D.}\ \bibnamefont {Stone}}, \ and\
  \bibinfo {author} {\bibfnamefont {L.}~\bibnamefont {Yang}},\ }\bibfield
  {title} {\enquote {\bibinfo {title} {Coherent perfect absorption at an
  exceptional point},}\ }\href {\doibase 10.1126/science.abj1028} {\bibfield
  {journal} {\bibinfo  {journal} {Science}\ }\textbf {\bibinfo {volume}
  {373}},\ \bibinfo {pages} {1261--1265} (\bibinfo {year} {2021})}\BibitemShut
  {NoStop}%
\bibitem [{\citenamefont {Goldzak}\ \emph {et~al.}(2018)\citenamefont
  {Goldzak}, \citenamefont {Mailybaev},\ and\ \citenamefont
  {Moiseyev}}]{Goldzak18}%
  \BibitemOpen
  \bibfield  {author} {\bibinfo {author} {\bibfnamefont {T.}~\bibnamefont
  {Goldzak}}, \bibinfo {author} {\bibfnamefont {A.~A.}\ \bibnamefont
  {Mailybaev}}, \ and\ \bibinfo {author} {\bibfnamefont {N.}~\bibnamefont
  {Moiseyev}},\ }\bibfield  {title} {\enquote {\bibinfo {title} {Light stops at
  exceptional points},}\ }\href {\doibase 10.1103/PhysRevLett.120.013901}
  {\bibfield  {journal} {\bibinfo  {journal} {Phys. Rev. Lett.}\ }\textbf
  {\bibinfo {volume} {120}},\ \bibinfo {pages} {013901} (\bibinfo {year}
  {2018})}\BibitemShut {NoStop}%
\bibitem [{\citenamefont {Laha}\ \emph {et~al.}(2020)\citenamefont {Laha},
  \citenamefont {Dey}, \citenamefont {Gandhi}, \citenamefont {Biswas},\ and\
  \citenamefont {Ghosh}}]{Laha20}%
  \BibitemOpen
  \bibfield  {author} {\bibinfo {author} {\bibfnamefont {A.}~\bibnamefont
  {Laha}}, \bibinfo {author} {\bibfnamefont {S.}~\bibnamefont {Dey}}, \bibinfo
  {author} {\bibfnamefont {H.~K.}\ \bibnamefont {Gandhi}}, \bibinfo {author}
  {\bibfnamefont {A.}~\bibnamefont {Biswas}}, \ and\ \bibinfo {author}
  {\bibfnamefont {S.}~\bibnamefont {Ghosh}},\ }\bibfield  {title} {\enquote
  {\bibinfo {title} {Exceptional point and toward mode-selective optical
  isolation},}\ }\href {\doibase 10.1021/acsphotonics.9b01646} {\bibfield
  {journal} {\bibinfo  {journal} {ACS Photonics}\ }\textbf {\bibinfo {volume}
  {7}},\ \bibinfo {pages} {967--974} (\bibinfo {year} {2020})}\BibitemShut
  {NoStop}%
\bibitem [{\citenamefont {Wiersig}(2020)}]{Wiersig20}%
  \BibitemOpen
  \bibfield  {author} {\bibinfo {author} {\bibfnamefont {J.}~\bibnamefont
  {Wiersig}},\ }\bibfield  {title} {\enquote {\bibinfo {title} {Review of
  exceptional point-based sensors},}\ }\href {\doibase 10.1364/PRJ.396115}
  {\bibfield  {journal} {\bibinfo  {journal} {Photon. Res.}\ }\textbf {\bibinfo
  {volume} {8}},\ \bibinfo {pages} {1457--1467} (\bibinfo {year}
  {2020})}\BibitemShut {NoStop}%
\bibitem [{\citenamefont {Naghiloo}\ \emph {et~al.}(2019)\citenamefont
  {Naghiloo}, \citenamefont {Abbasi}, \citenamefont {Joglekar},\ and\
  \citenamefont {Murch}}]{Naghiloo2019}%
  \BibitemOpen
  \bibfield  {author} {\bibinfo {author} {\bibfnamefont {M.}~\bibnamefont
  {Naghiloo}}, \bibinfo {author} {\bibfnamefont {M.}~\bibnamefont {Abbasi}},
  \bibinfo {author} {\bibfnamefont {Y.~N.}\ \bibnamefont {Joglekar}}, \ and\
  \bibinfo {author} {\bibfnamefont {K.~W.}\ \bibnamefont {Murch}},\ }\bibfield
  {title} {\enquote {\bibinfo {title} {Quantum state tomography across the
  exceptional point in a single dissipative qubit},}\ }\href {\doibase
  10.1038/s41567-019-0652-z} {\bibfield  {journal} {\bibinfo  {journal} {Nature
  Physics}\ }\textbf {\bibinfo {volume} {15}},\ \bibinfo {pages} {1232--1236}
  (\bibinfo {year} {2019})}\BibitemShut {NoStop}%
\bibitem [{\citenamefont {Minganti}\ \emph {et~al.}(2019)\citenamefont
  {Minganti}, \citenamefont {Miranowicz}, \citenamefont {Chhajlany},\ and\
  \citenamefont {Nori}}]{Minganti2019}%
  \BibitemOpen
  \bibfield  {author} {\bibinfo {author} {\bibfnamefont {F.}~\bibnamefont
  {Minganti}}, \bibinfo {author} {\bibfnamefont {A.}~\bibnamefont
  {Miranowicz}}, \bibinfo {author} {\bibfnamefont {R.~W.}\ \bibnamefont
  {Chhajlany}}, \ and\ \bibinfo {author} {\bibfnamefont {F.}~\bibnamefont
  {Nori}},\ }\bibfield  {title} {\enquote {\bibinfo {title} {Quantum
  exceptional points of non-hermitian hamiltonians and liouvillians: The
  effects of quantum jumps},}\ }\href {\doibase 10.1103/PhysRevA.100.062131}
  {\bibfield  {journal} {\bibinfo  {journal} {Phys. Rev. A}\ }\textbf {\bibinfo
  {volume} {100}},\ \bibinfo {pages} {062131} (\bibinfo {year}
  {2019})}\BibitemShut {NoStop}%
\bibitem [{\citenamefont {Ge}\ and\ \citenamefont {T\"ureci}(2013)}]{Ge13APT}%
  \BibitemOpen
  \bibfield  {author} {\bibinfo {author} {\bibfnamefont {L.}~\bibnamefont
  {Ge}}\ and\ \bibinfo {author} {\bibfnamefont {H.~E.}\ \bibnamefont
  {T\"ureci}},\ }\bibfield  {title} {\enquote {\bibinfo {title} {Antisymmetric
  $\mathcal{PT}$-photonic structures with balanced positive- and negative-index
  materials},}\ }\href {\doibase 10.1103/PhysRevA.88.053810} {\bibfield
  {journal} {\bibinfo  {journal} {Phys. Rev. A}\ }\textbf {\bibinfo {volume}
  {88}},\ \bibinfo {pages} {053810} (\bibinfo {year} {2013})}\BibitemShut
  {NoStop}%
\bibitem [{\citenamefont {Zhang}\ \emph {et~al.}(2019)\citenamefont {Zhang},
  \citenamefont {Jiang},\ and\ \citenamefont {Chan}}]{Zhang2019APTEP}%
  \BibitemOpen
  \bibfield  {author} {\bibinfo {author} {\bibfnamefont {X.-L.}\ \bibnamefont
  {Zhang}}, \bibinfo {author} {\bibfnamefont {T.}~\bibnamefont {Jiang}}, \ and\
  \bibinfo {author} {\bibfnamefont {C.~T.}\ \bibnamefont {Chan}},\ }\bibfield
  {title} {\enquote {\bibinfo {title} {Dynamically encircling an exceptional
  point in anti-parity-time symmetric systems: asymmetric mode switching for
  symmetry-broken modes},}\ }\href {\doibase 10.1038/s41377-019-0200-8}
  {\bibfield  {journal} {\bibinfo  {journal} {Light: Science {\&}
  Applications}\ }\textbf {\bibinfo {volume} {8}},\ \bibinfo {pages} {88}
  (\bibinfo {year} {2019})}\BibitemShut {NoStop}%
\bibitem [{\citenamefont {Feng}\ and\ \citenamefont {Sun}(2022)}]{Feng22APTEP}%
  \BibitemOpen
  \bibfield  {author} {\bibinfo {author} {\bibfnamefont {Z.}~\bibnamefont
  {Feng}}\ and\ \bibinfo {author} {\bibfnamefont {X.}~\bibnamefont {Sun}},\
  }\bibfield  {title} {\enquote {\bibinfo {title} {Harnessing dynamical
  encircling of an exceptional point in anti-$\mathcal{P}\mathcal{T}$-symmetric
  integrated photonic systems},}\ }\href {\doibase
  10.1103/PhysRevLett.129.273601} {\bibfield  {journal} {\bibinfo  {journal}
  {Phys. Rev. Lett.}\ }\textbf {\bibinfo {volume} {129}},\ \bibinfo {pages}
  {273601} (\bibinfo {year} {2022})}\BibitemShut {NoStop}%
\bibitem [{\citenamefont {Qi}\ \emph {et~al.}(2021)\citenamefont {Qi},
  \citenamefont {Hu}, \citenamefont {Wang},\ and\ \citenamefont
  {Gong}}]{Qi2021APTEP}%
  \BibitemOpen
  \bibfield  {author} {\bibinfo {author} {\bibfnamefont {H.}~\bibnamefont
  {Qi}}, \bibinfo {author} {\bibfnamefont {X.}~\bibnamefont {Hu}}, \bibinfo
  {author} {\bibfnamefont {X.}~\bibnamefont {Wang}}, \ and\ \bibinfo {author}
  {\bibfnamefont {Q.}~\bibnamefont {Gong}},\ }\bibfield  {title} {\enquote
  {\bibinfo {title} {Encircling an exceptional point in a multiwaveguide
  anti--parity-time-symmetry system},}\ }\href {\doibase
  10.1103/PhysRevA.103.063520} {\bibfield  {journal} {\bibinfo  {journal}
  {Phys. Rev. A}\ }\textbf {\bibinfo {volume} {103}},\ \bibinfo {pages}
  {063520} (\bibinfo {year} {2021})}\BibitemShut {NoStop}%
\bibitem [{\citenamefont {Ge}\ \emph {et~al.}(2011)\citenamefont {Ge},
  \citenamefont {Chong}, \citenamefont {Rotter}, \citenamefont {T\"ureci},\
  and\ \citenamefont {Stone}}]{Ge11}%
  \BibitemOpen
  \bibfield  {author} {\bibinfo {author} {\bibfnamefont {L.}~\bibnamefont
  {Ge}}, \bibinfo {author} {\bibfnamefont {Y.~D.}\ \bibnamefont {Chong}},
  \bibinfo {author} {\bibfnamefont {S.}~\bibnamefont {Rotter}}, \bibinfo
  {author} {\bibfnamefont {H.~E.}\ \bibnamefont {T\"ureci}}, \ and\ \bibinfo
  {author} {\bibfnamefont {A.~D.}\ \bibnamefont {Stone}},\ }\bibfield  {title}
  {\enquote {\bibinfo {title} {Unconventional modes in lasers with spatially
  varying gain and loss},}\ }\href {\doibase 10.1103/PhysRevA.84.023820}
  {\bibfield  {journal} {\bibinfo  {journal} {Phys. Rev. A}\ }\textbf {\bibinfo
  {volume} {84}},\ \bibinfo {pages} {023820} (\bibinfo {year}
  {2011})}\BibitemShut {NoStop}%
\end{thebibliography}%

\end{document}